\documentclass[aps,prb,reprint,superscriptaddress,showpacs,showkeys,floatfix]{revtex4-2}
\usepackage{graphicx}
\usepackage{amsmath}
\usepackage{xcolor}
\usepackage{natbib}
\newcommand{\uu}[1]{\ensuremath{\,{\mathrm{#1}}}}

\begin{document}


\title{Graphene on silicon: effects of the silicon surface orientation on the work function and carrier density of graphene}


\author{Y. W. Sun}
\email{yiwei.sun@qmul.ac.uk}
\affiliation{School of Engineering and Materials Science, Queen Mary University of London, London E1 4NS, United Kingdom}
\author{D. Holec}
\email{david.holec@unileoben.ac.at}
\affiliation{Department of Materials Science, Montanuniversit\"{a}t Leoben, Leoben 8700, Austria}
\author{D. Gehringer}
\affiliation{Department of Materials Science, Montanuniversit\"{a}t Leoben, Leoben 8700, Austria}
\author{L. Li}
\affiliation{College of Information Science and Electronic Engineering, Zhejiang University, Hangzhou 310027, China}
\author{O. Fenwick}
\affiliation{School of Engineering and Materials Science, Queen Mary University of London, London E1 4NS, United Kingdom}
\author{D. J. Dunstan}
\affiliation{School of Physics and Astronomy, Queen Mary University of London, London E1 4NS, United Kingdom}
\author{C. J. Humphreys}
\email{c.humphreys@qmul.ac.uk}
\affiliation{School of Engineering and Materials Science, Queen Mary University of London, London E1 4NS, United Kingdom}

\date{\today}

\begin{abstract}
Density functional theory has been employed to study graphene on the (111), (100) and (110) surfaces of silicon (Si) substrates. There are several interesting findings. First, carbon atoms in graphene form covalent bonds with Si atoms, when placed close enough on Si (111) and (100) surfaces, but not on the (110) surface. The presence of a Si (111) surface shifts the Fermi level of graphene into its conduction band, resulting in an increase of the work function by 0.29 eV and of the electron density by three orders of magnitude. The carrier density of graphene can also be increased by eighty times on a Si (100) substrate without doping, due to the modification of the density of states near the Dirac point. No interfacial covalent bond can be formed on Si (110). These striking effects that different orientations of a silicon substrate can have on the properties of graphene are related to the surface density of the silicon surface. Applying the results to a real device of a specific orientation requires further consideration of surface reconstructions, lattice mismatch, temperature, and environmental effects.
\end{abstract}


\maketitle
\section{Introduction}
The extraordinary properties of graphene reported in the literature are almost always recorded on a substrate.\cite{Novoselov04} Substrates can be crystals, such as Si and sapphire, and they can have different surface orientations. The properties of monolayer graphene are expected to depend upon the orientation of the substrate surface. We have therefore investigated theoretically the interface structure of graphene on three different Si surfaces: (100), (110) and (111). In particular, we have studied the spacing between the graphene and the silicon surface, and whether the carbon and silicon atoms form covalent bonds.

The nature of the interface affects the work function and the carrier density of graphene, which are key properties for electronic and photoelectric devices.\cite{Zhu09,Hwang07} Calculations reported the work function of freestanding graphene in vacuum to be around 4.5 eV.\cite{Khomyakov09} Experimentally the work function of graphene has been measured on SiO$_2$ and the values vary from 4.6 to 5.2 eV.\cite{Yan12,Song12,Song13,Park11} The increase was attributed to hydroxyl groups at the SiO$_2$ surface, presumably from atmospheric water vapour.\cite{Vives16,Song15} These values further vary with different metal contacts.\cite{Song12} It is common to dope graphene to tune its work function for specific applications,\cite{Kim10,Ryu10} such as increasing the power conversion efficiency of a graphene-Si solar cell.\cite{Xu16} In this paper, we quantify the increase of the work function of graphene from the mere presence of Si substrates of various surface orientations.

The carrier density of pristine undoped graphene is low as the density of states (DOS) around the Fermi level (at the Dirac point) is low.\cite{Novoselov04} The theoretical value is about $10^{11}$ cm$^{-2}$ (about 10$^{18}$ cm$^{-3}$ considering graphene to be 3.4 \AA{} thick \cite{Tuinstra70}, as expected for a semi-metal) at 300 K,\cite{Fang07} compared to $9.65\times10^9$ cm$^{-3}$ for Si,\cite{Altermatt03} $2.33\times10^{13}$ cm$^{-3}$ for germanium (semiconductors),\cite{Madelung} and $8.49\times10^{22}$ cm$^{-3}$ for copper (metal) from the simple Drude model. These theoretical values are generally consistent with experimental measurements. For example, the experimental value for intrinsic Si is $1.0\times10^{10}$ cm$^{-3}$.\cite{Altermatt03} However, exfoliated monolayer graphene on a SiO$_2$ substrate has a higher carrier density of about $10^{13}$ cm$^{-2}$.\cite{Yin14} Here we quantify how much a Si substrate can increase the carrier density of graphene, depending on its surface orientations.

This paper quantifies the impact of different surface orientations of Si substrates on graphene and demonstrates the physics behind the effects. 

\section{Methods}
We modelled monolayer graphene on a Si substrate with three different surface orientations, (111), (100) and (110). The simulation box contained 18 C atoms and 36 Si atoms for graphene on Si (111), 20 C atoms and 52 Si atoms for graphene on Si (100), and 30 C atoms and 75 Si atoms for graphene on Si (110). The bottom layers of Si atoms were fixed at the Si lattice constant and the top layers were relaxed. For example, for Si (111), we fixed two and a half double-layers at the bottom, and relaxed the top two double-layers. The in-plane dimensions of the simulation box were kept constant. The surface is not large enough to undergo the usual ($7\times7$) reconstructions, and no other surface reconstruction is seen.

Graphene was strained to the Si lattice on each surface to meet the in-plane periodic boundary condition. For example, graphene is at 4.6\% isotropic tensile strain on Si (111). Anisotropic tensile strain is introduced on other surfaces. We compensate for the effects of the in-plane strain by comparing the graphene on the Si surface with graphene similarly strained but free-standing. In the initial stacking, some C atoms were directly on top of some Si atoms. The structures were later relaxed while keeping in-plane lattice parameters fixed (thus simulating a thick Si substrate). On the Si (111) surface, where the relaxed structure possesses high (and likely artificial) symmetry, five more random initial stackings (i.e., random initial in-plane displacement of the graphene sheet) were introduced.

We wish to investigate if stable covalent bonds can be formed when the graphene is placed, or deposited, on a Si surface. We note that the separation of van der Waals (vdW) bonded graphene layers in graphite is 3.4 \AA{} and the bond length of SiC in the very stiff 6H-SiC is 1.9 \AA{}.\cite{Zetterling} On each surface orientation we therefore relaxed our structure from two different initial graphene-Si separations: 4.0 \AA{} and 1.5 \AA{}. The graphene-Si separation is defined as follows. The position of a graphene plane was defined by the centres of the four C atoms at corners of the plane in a simulation box. The distance from this plane to the centres of the topmost Si atoms is defined as the graphene-Si separation, as shown in Fig.~\ref{Geo}. 

The structures and properties of these systems were found by density functional theory (DFT) \cite{DFT1,DFT2} using the Vienna Ab-initio Simulation Package (VASP) \cite{VASP}. We used the generalised gradient approximation (GGA), parameterized by Perdew, Burke and Ernzerhof \cite{GGA} for the exchange-correlation and the projector augmented-wave method pseudopotentials \cite{PP} for carbon. The calculated total energy excludes contributions from the core electrons. 
In this paper we present the calculated energy relative to the Fermi level ($E-E_{F}$), and we interpret the difference between the Fermi and vacuum level as the work function. The effects of vdW interactions were included using the Grimme method \cite{vdW} as implemented in the VASP code. The structural models were visualised using the VESTA software \cite{vesta}.

\section{Results and Discussions}
\subsection{Interface structure}
\begin{figure*}
\includegraphics[width=\textwidth]{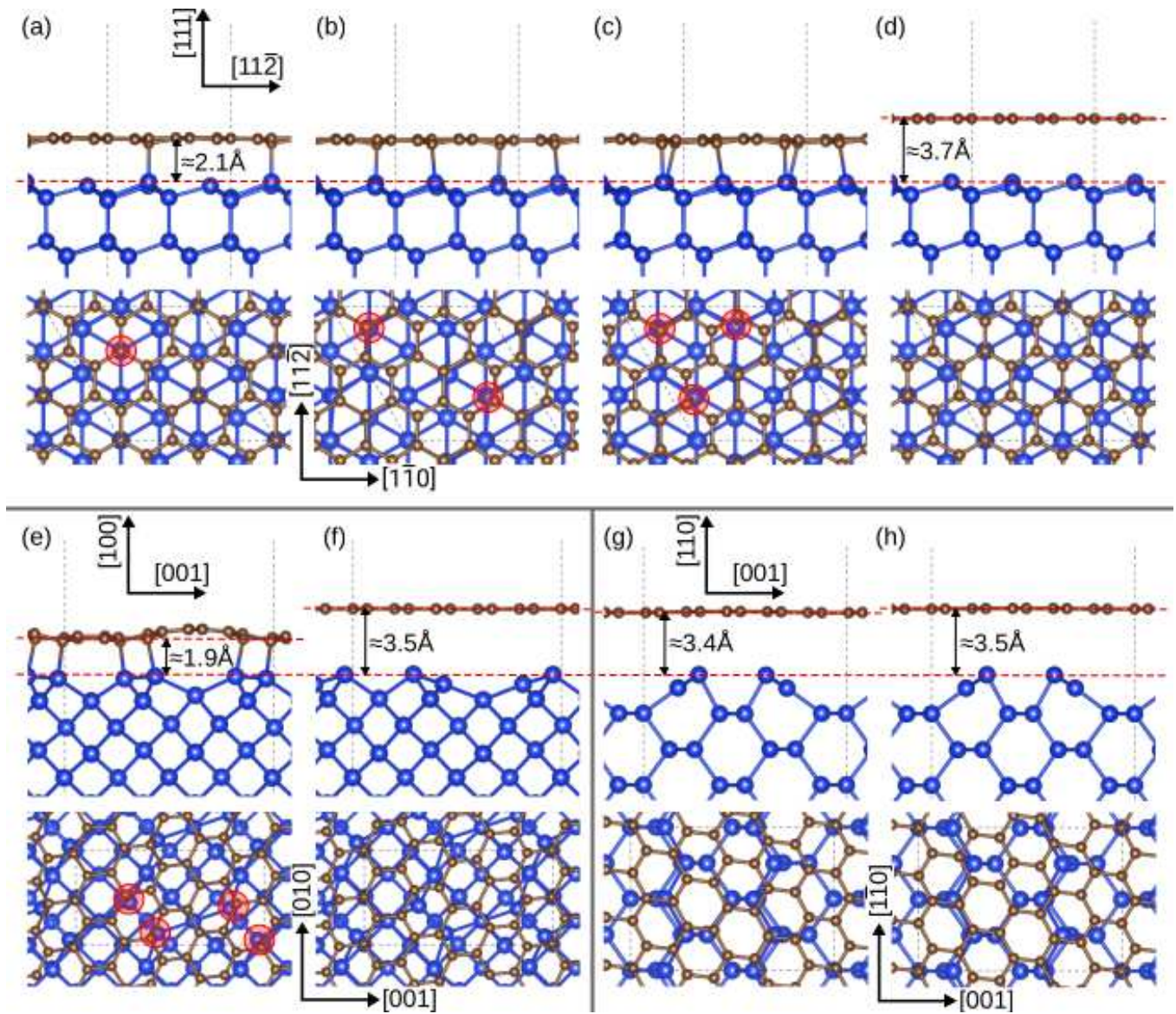}
\caption{The relaxed structure of monolayer graphene on various Si surfaces from different initial separations, viewed along the $a$-axis (side view), with the corresponding plan view along the $c$-axis beneath (which covers a slightly smaller range). (a) Si (111), 1.5 \AA; (b) Si (111), 1.5 \AA, a different stacking order; (c) Si (111), 1.5 \AA, another different stacking order; (d) Si (111), 4.0 \AA; (e) Si (100), 1.5 \AA; (f) Si (100), 4.0 \AA; (g) Si (110), 1.5 \AA; (h) Si (110), 4.0 \AA. The simulation boxes are labeled by dashed lines. The Si and C atoms forming bonds are labeled by red circles in the plan view.}
\label{Geo}
\end{figure*}

Fig.~\ref{Geo} shows the side view of the relaxed structure of graphene on Si (111) ((a)--(d)), (100) ((e) and (f)) and (110) ((g) and (h)) surfaces, with the corresponding plan view beneath, from two different initial separations, 1.5 ((a), (e) and (g)) and 4.0 \AA{} ((d), (f) and (h)). Note that the relaxed structure possesses high symmetry on Si (111) (the plan view in Fig.~\ref{Geo} (a)). We relaxed the system after several random in-plane displacements of the graphene sheet and obtained two more (meta-)stable states ((b) and (c)). From Fig.~\ref{Geo} (d), (f) and (h), when initially placed far away, a graphene layer can be vdW bonded to all the three orientations of the Si surface, with no visible out-of-plane perturbation (bulging) on the $sp^2$-network of graphene. The distance between the graphene plane and the top Si atoms varies between $\approx3.5$ and $\approx3.7\uu{\AA}$. The difference in the distance here can be attributed to the different Si surfaces and the way we define the positions of these surfaces. The Si (111) surface consists of a double-layer (Fig.~\ref{Geo} (d)), which is denser and effectively more robust (as the surface is hardly disrupted by the graphene layer above), resulting in a larger interlayer distance with graphene than for the (100) (Fig.~\ref{Geo} (f)) and (110) (Fig.~\ref{Geo} (h)) surfaces. The larger vdW distance between graphene and Si than between graphene layers in graphite (3.34 \AA{} \cite{Tuinstra70}) indicates a weaker vdW attraction, or that the dangling bonds of Si extend further than the $\pi$-orbitals of graphene.

	The interesting finding here is that when the initial separation between a graphene layer and the Si surface was set to be 1.5\uu{\AA}, some Si and C atoms approach each other, forming Si-C bonds on the (111) (Fig.~\ref{Geo} (a)--(c)) and (100) surfaces (Fig.~\ref{Geo} (e)), but not on the (110) surface (Fig.~\ref{Geo} (g)). On the Si (111) surface, 1 bond per simulation box area is formed (corresponding to a Si-C bond density of 0.018 bond/\AA$^2$, Fig. \ref{Geo} (a)), however, 4 Si-C bonds per simulation box area are formed on the (100) surface (bond density of 0.067 bond/\AA$^2$, Fig.~\ref{Geo} (e)). As a result of more bonds, the interlayer spacing on Si (100) ($\approx1.9\uu{\AA}$) is smaller than that on Si (111) ($\approx2.1\uu{\AA}$). In contrast, on the (110) surface, graphene relaxes to a distance of $\approx3.4$ \AA{}, even from 1.5\uu{\AA}, and remains visibly flat (Fig.~\ref{Geo} (g)). No Si-C bond is formed on the Si (110) surface. The different vdW distances on the (110) surface from the different initial separations correspond to a difference in energy of 0.07 meV/atom, below the accuracy of the present calculations.

The relaxed structure of graphene on Si (111) possesses high (and likely artificial) symmetry: one C atom is directly above one Si atom, and the whole model exhibits a 3-fold rotational symmetry along the vertical axis through these vertically overlapping C and Si atoms (Fig.~\ref{Geo} (a)). To understand the effect of stacking orders on the formation of Si-C bonds, we relaxed the structure from additional initial stackings by giving random in-plane translational displacements to the graphene plane within the unit cell dimensions from the high-symmetry structure. With the initial separations at $1.5\uu{\AA}$, two more (meta)stable states were obtained: one with 2 bonds per simulation box area (density of bonds $0.039\uu{bond/\AA^2}$, Fig.~\ref{Geo} (b)) and the other with 3 bonds per box area ($0.058\uu{bond/\AA^2}$, Fig.~\ref{Geo} (c)). These two additional bonded states on Si (111) are very similar in atomic positions despite different densities of bonds. The relaxations starting at the large separation of $4\uu{\AA}$, but of different initial stackings, always converged to the same state of Fig.~\ref{Geo} (d).

We propose a possible interpretation for the formation of these interlayer Si-C bonds: it is related to the surface density of the Si substrates. The Si (111) surfaces consists of a double plane, and is of the highest density. Hence, there is a higher probability of having a Si atom vertically close enough to a C atom, to form a bond. Si (100) plane is of the lowest density, but because of that, atoms at the surface are easily displaced from their original positions and the top two layers merge into one after a graphene layer is placed on top. After the merge, the surface density doubles and moreover, compared to the (111) plane, the surface Si atoms are still quite free to move, further promoting the formation of the bonds. For the (110) plane, its surface is also disrupted by the graphene on top, similarly to the (100) plane, but the low surface density further decreases, reducing the chance of forming Si-C bonds.

\subsection{Interlayer Potential}
\begin{figure}
    {\sffamily (a) \hfill (111) surface \hfill \mbox{}}\smallskip\\
	\includegraphics[width=0.9\columnwidth]{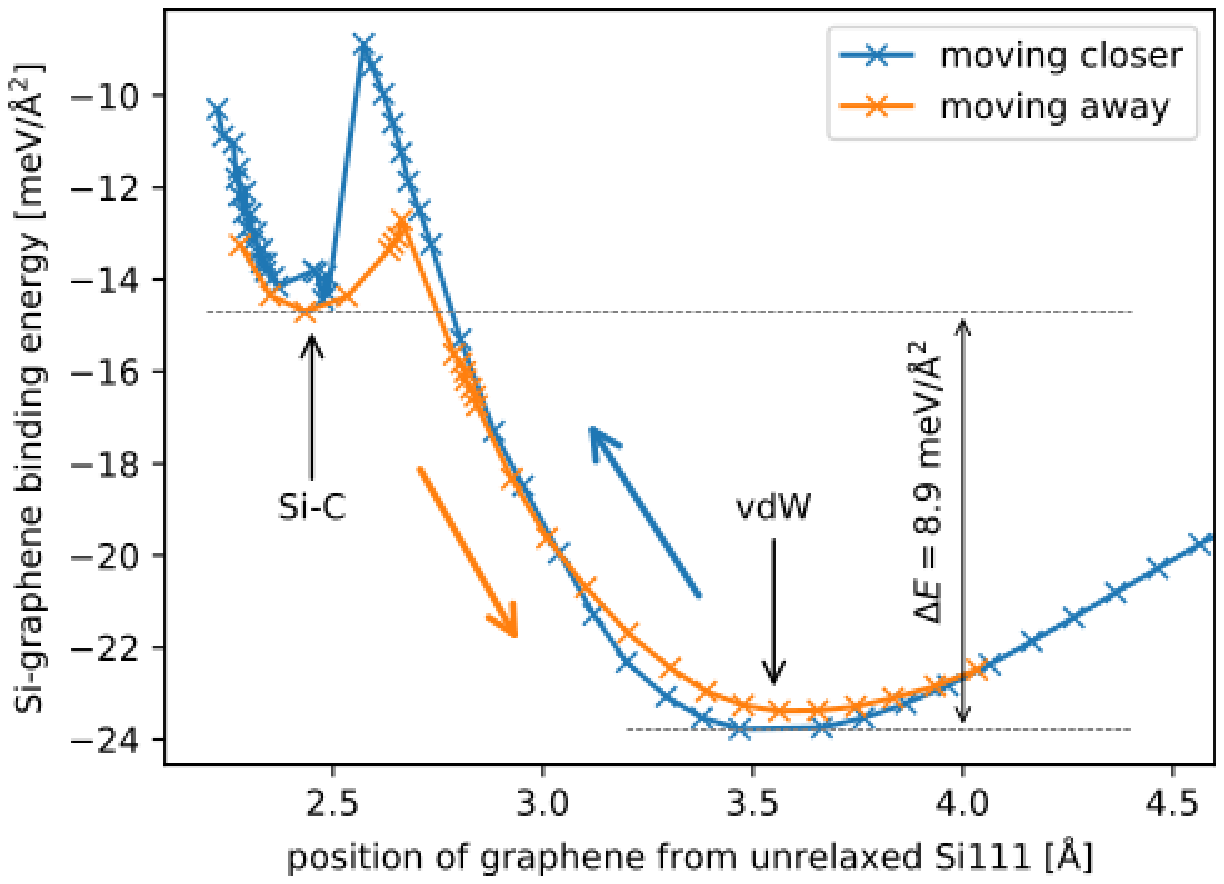}\medskip\\
	{\sffamily (b) \hfill (100) surface \hfill \mbox{}}\smallskip\\
	\includegraphics[width=0.9\columnwidth]{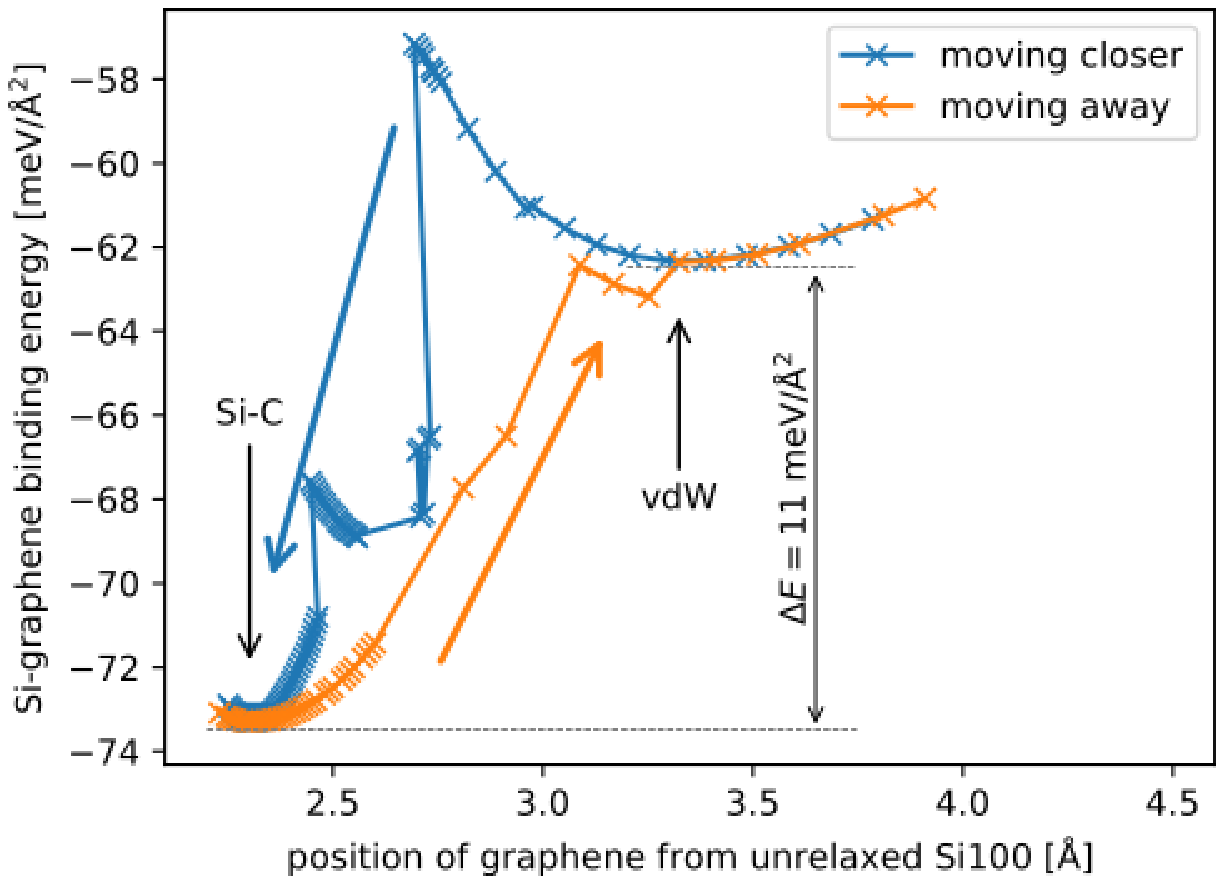}
	\caption{The binding energy (per in-plane unit area) is plotted against the separation between the graphene and Si surface, on (a) (111) and (b) (100), while displacing the graphene plane towards (blue) and away from (orange) the Si surface.}
	\label{el}
\end{figure}

Plotting the interlayer potential energy (binding energy) against interlayer distance (Fig.~\ref{el}) provides further insights into the significantly different structures of graphene on the different substrates. On the (110) surface, there are no (meta)stable states other than the vdW-bonded state. We investigate the differences in energy and the heights of barriers between (meta)stable states on the (111) and (100) surfaces. We calculated the energy when displacing the graphene layer along the $c$-axis towards, and away from a Si surface. We plotted the interlayer binding energy against the interlayer distance in Fig.~\ref{el}. At each fixed $z$-coordinates of the corner four C atoms and the fixed positions of the bottom Si layers (note that the plotted distance is to the topmost Si atoms before relaxation), we relaxed the graphene and the top Si layers, and calculated the energy of the relaxed state. From Fig.~\ref{el}, it is clear that the vdW bonded state is the stable state for graphene on Si (111) and the covalently bonded state is the stable state on Si (100).

In addition to the difference in binding energy between the states, see Fig.~\ref{el}, we can also obtain the energy required to form interfacial Si-C bonds relative to the energy at the vdW distance, and that required to break the formed bonds. On the (111) surface, forming Si-C bonds requires 15 meV/\AA{}$^2$ and breaking these bonds takes only 1.5 meV/\AA{}$^2$. On the (100) surface, it takes 5 meV/\AA{}$^2$ to form bonds and 11 meV/\AA{}$^2$ to break them. On both surfaces, there is a clear hysteresis when moving graphene towards and away from Si. In particular, on the (100) surface, two Si-C bonds (per simulation box) form first before the global minimum energy with four bonds is reached, but there is no intermediate state predicted when breaking these bonds.

\subsection{Work Function}
\begin{figure*}
	\includegraphics[width=1.8\columnwidth]{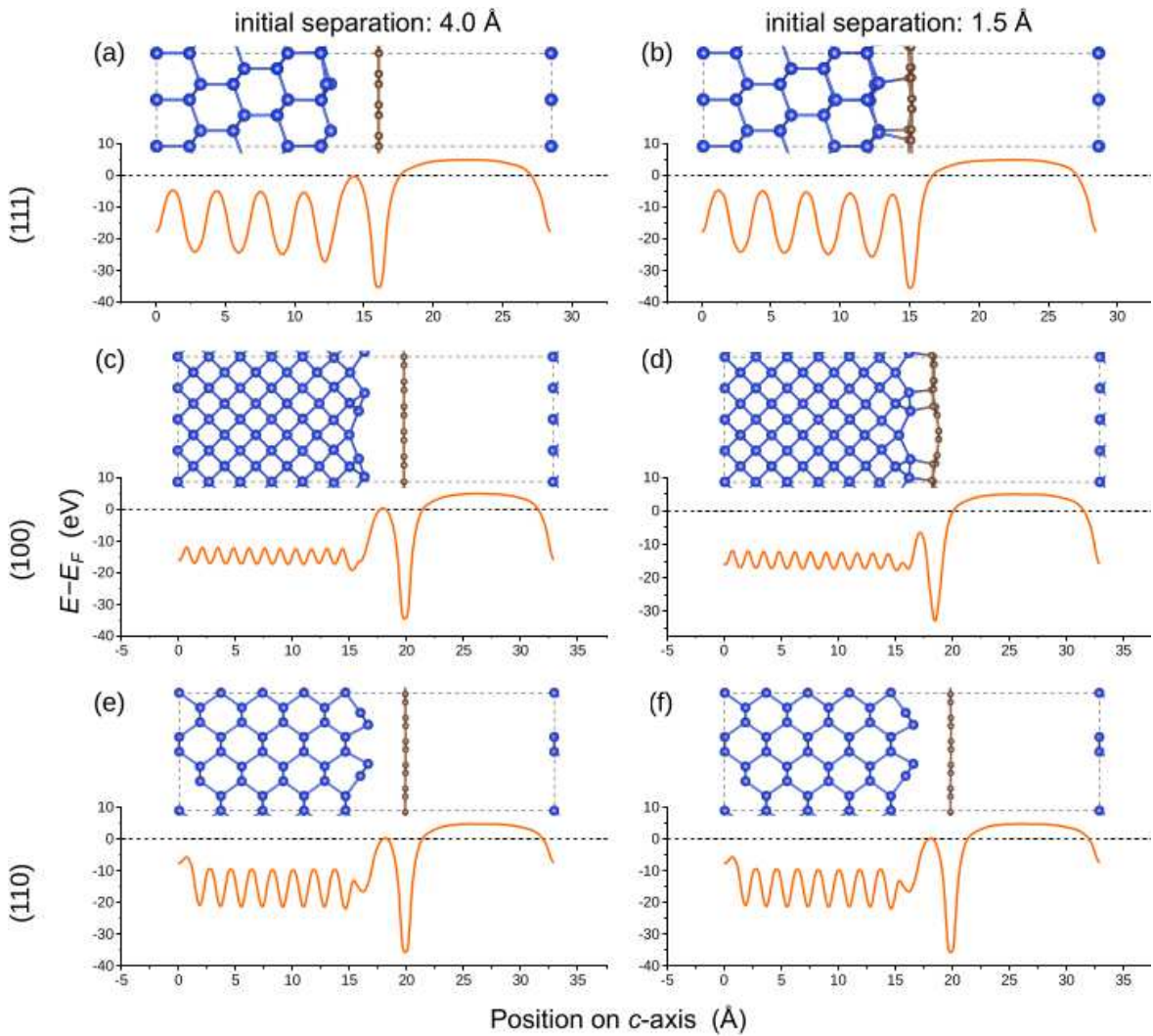}
	\caption{The difference between the electrostatic potential and the Fermi level is plotted along the $c$-axis perpendicular to the graphene plane, for graphene on (a) Si (111) with initial separation of 4.0 \AA, (b) Si (111) with 1.5 \AA, (c) Si (100) with 4.0 \AA, (d) Si (100) with 1.5 \AA, (e) Si (110) with 4.0 \AA, and (f) Si (110) with 1.5 \AA. The structures are shown above. The plateau in the energy corresponds to the vacuum between graphene and the first Si layer in the next unit cell. The Fermi level (the zero) is shown by a dashed line.}
	\label{esp}
\end{figure*}

We calculated the Fermi level of the system of graphene on Si and plotted the difference between the electrostatic potential energy and the Fermi level along the $c$-axis perpendicular to the graphene plane in Fig.~\ref{esp}. The structures are shown above each plot. The potential well for the graphene layer is deeper than for the Si layers because there are more C atoms per unit area in a graphene plane than Si atoms in a Si layer. For example, in our simulation unit cell, there are 8 Si atoms in a Si (111) double-layer and 18 C atoms in a graphene plane. The plateau on the right of the graphene corresponds to the vacuum level, extending to the first Si layer of the next unit cell. The Fermi level (as the zero of energy) is shown by dashed lines. The work function is the difference between the Fermi level and the electrostatic potential energy in the vacuum. Table~\ref{t1} lists the work function values. As mentioned above, the graphene is strained to match the Si lattices, to meet the in-plane periodic boundary condition. The strain modifies the work function of graphene, comparing strained vs. pristine graphene in Table~\ref{t1}. We compare the work function of graphene strained to a specific Si surface plane, with and without the presence of the actual Si substrate beneath.

\begin{table}
	\caption{The difference between the calculated Fermi level and electrostatic potential energy at vacuum, which we interpret as the work function are listed for unstrained pristine graphene, graphene strained to Si surfaces but unsupported, and strained graphene with the corresponding Si surface beneath. On each surface, there are two structures relaxed from two different initial separations between graphene and Si.}
	\label{t1}
	\begin{tabular}{l | c c c}
		\hline\hline
		Si surface	&Work function \\
			&(eV) \\
		\hline
		Pristine graphene	&4.22 \\
		Strained to Si (111)	&4.60 \\
		On Si (111) 4.0 \AA		&4.86 \\
		On Si (111) 1.5 \AA 	&4.89	\\
	Strained to Si (100) 	&4.81 \\
	On Si (100) 4.0 \AA		&4.92 \\
	On Si (100) 1.5 \AA 	&5.02	\\
	Strained to Si (110) 	&4.62 \\
	On Si (110) 4.0 \AA		&4.78 \\
	On Si (110) 1.5 \AA 	&4.79	\\
		\hline\hline	
	\end{tabular}
\end{table}

There are two caveats to mention before we discuss these results. First, the in-plane strain increases the work function significantly (as much as 0.59 eV when strained to Si (100)) compared with unstrained graphene, but here we focus on the further increase by the presence of Si substrates. Second, because of the anisotropic in-plane strain of graphene to match the Si lattice, there are small band gaps induced in graphene, which is artificial and of no interest to this paper.

The increase of the work function of graphene on Si surfaces at their vdW distances to the graphene planes is 0.27, 0.11 and 0.16 eV on Si (111), (100) and (110) surfaces respectively. The change of the graphene work function can be related to a surface dipole, induced by an asymmetric distribution of electronic orbitals on each side of a graphene layer when placed on a Si surface. The different increases on different Si surface orientations is due to the different electrostatic potentials at the interface of graphene and Si (as shown in Fig. \ref{esp} (a), (c), (e)), to which the work function is sensitive.\cite{Bardeen47} The different local electric fields are a result of different interface structures. The Si (111) surface layer is similar to the layers in the bulk. The top two single-layers at the Si (100) surface merge into one, whilst the surface layer of Si (110) has only about half as many atoms as those in the bulk.

Charge transfer is a common cause for changes of the graphene work function. An important result here is that the Fermi level is very close to the Dirac point of graphene on Si (100) and (110) surfaces, but on the Si (111) surface the Fermi level is shifted a long way, about 0.7 eV, into the conduction band. The Fermi level of the system is higher than it is in either graphene or Si alone. This unexpected result is due to the significant modification of the DOS of graphene, which will be quantified and interpreted in the next section after the DOS results are presented. The consequential increase in the work function of graphene on Si (111) is 0.27 eV, the largest among all the cases.

The work function of graphene is affected most on the most robust (atomically dense) surface (111) and least on the surface (100), where the atoms are free to move. The robustness can be related to the surface density for the three surfaces in this paper. We propose a possible explanation for the effect of the surface orientation. When two very robust surfaces (\textit{e.g.} graphene and Si (111)) contact, nuclei are held in position by the stable in-plane network of each surface and there will be little disruption on the in-plane structures. With stable nuclear positions, electronic orbitals (especially those out-of-plane, such as the $\pi$-orbitals of graphene) may significantly redistribute to lower the energy when the graphene approaches a Si surface, inducing surface dipoles or modifying existing ones. 

Finally, we consider the effect of graphene forming covalent bonds with Si substrates on the work function of graphene. The work function is increased by 0.02 and 0.11 eV, respectively, on Si (111) and (100) surfaces when covalent bonds are formed, compared to the vdW-bonded states. Forming covalent bonds is a lesser impact than the mere presence of a Si substrate on Si (111), which increases the work function by 0.27 eV. These two have an equal impact on the work function of graphene on Si (100).

\subsection{Carrier Density}
\begin{figure*}
	\includegraphics[width=1.8\columnwidth]{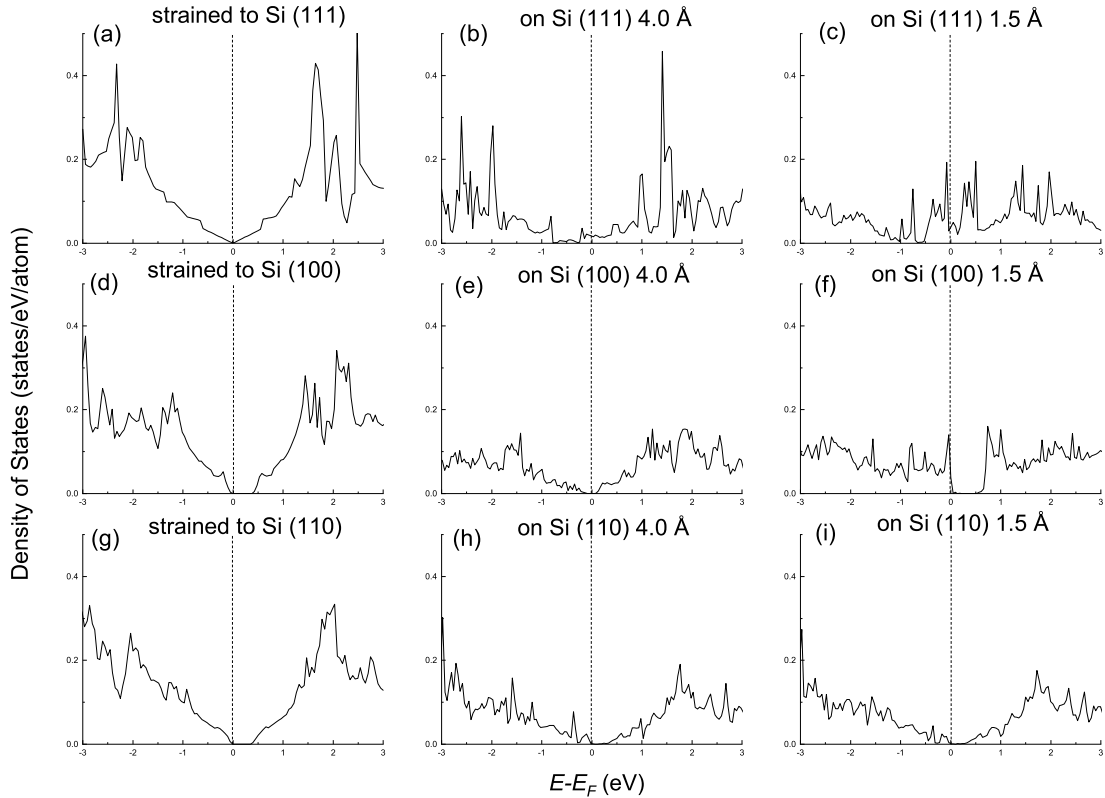}
	\caption{The densities of states (per atom, around the Fermi levels of the systems) of the carbon atoms are plotted, for graphene (a) strained to Si (111) but unsupported, (b) on Si (111) with an initial separation of 4.0 \AA, (c) on Si (111) with 1.5 \AA, (d) strained to Si (100) but unsupported, (e) on Si (100) with 4.0 \AA, (f) Si (100) with 1.5 \AA, (g) strained to Si (110) but unsupported, (h) Si (110) with 4.0 \AA, and (i) Si (110) with 1.5 \AA. Energy is referenced to the Fermi levels, which is shown by the chain-dotted lines at $E-E_F=0$.}
	\label{dos}
\end{figure*}

We obtain the electron (hole) density of graphene by integrating the product of the DOS of all the C atoms and the Fermi-Dirac distribution at 300 K, from the bottom (top) of the conduction (valence) band of graphene to the top (bottom). We choose the DOS of C atoms, because the transport of carriers in C is very different from Si (e.g. in mobility), and we are interested in carriers in the graphene, rather than in the bulk Si. The effect of temperature on the DOS of graphene (and Si) is very small.\cite{Yin14} Here we combine the DOS at 0 K with the Fermi Dirac distribution at 300 K, to demonstrate the effect of different Si surface orientations on the carrier density of graphene at room temperature. We present the DOS (near the Fermi level) of all the C atoms in each case in Fig. \ref{dos}. The Fermi levels (dashed lines) are at, or close to, the Dirac point (of zero DOS), except on the (111) surface (Fig.~\ref{dos} (b) and (c)).

Table~\ref{t2} lists the calculated carrier densities. Normally large changes in the carrier density are achieved by shifting the Fermi level, so the distance of the Fermi level from the edge of the valence or conduction band, whichever is closer, is also listed.
\begin{table}
	\caption{The calculated distance of the Fermi level from the edge of the valence or conduction band, whichever is closer, and the carrier density in the graphene are listed for unstrained pristine graphene, graphene strained to Si surfaces but unsupported, and strained graphene with a Si substrate of the corresponding surface orientation beneath. On each surface, there are two structures, relaxed from two different initial separations between graphene and Si. Electron density is indicated by $n$ and hole density by $p$.}\label{t2}
	\begin{tabular}{l | c c c}
		\hline\hline
		Si surface	& $E_{F}-E$ (edge of band)	&Carrier density \\
		&(eV)	&(cm$^{-2}$) \\
		\hline
		\\
		Pristine graphene	&0.042	&n=1$\times 10^{11}$ \\
		&&p=1$\times 10^{11}$ \\
		Strained to Si (111) &-0.011	&n=1$\times 10^{11}$ \\
		&&p=1$\times 10^{11}$ \\
		On Si (111) 4.0 \AA	&0.678	&n=2$\times 10^{13}$ \\
		On Si (111) 1.5 \AA &0.593	&n=1$\times 10^{14}$	\\
		Strained to Si (100) &0.025	&p=1$\times 10^{11}$ \\
		On Si (100) 4.0 \AA	&inside a band gap	&n=2$\times 10^{10}$ \\
		&&p=4$\times 10^{9}$ \\
		On Si (100) 1.5 \AA	&-0.034	&p=8$\times 10^{12}$ \\
		Strained to Si (110) &-0.004	&p=2$\times 10^{11}$ \\
		On Si (110) 4.0 \AA	&-0.003	&p=3$\times 10^{11}$ \\
		On Si (110) 1.5 \AA &0	&p=3$\times 10^{11}$	\\
		\hline\hline	
	\end{tabular}
\end{table}

The uncertainty of the data is as follows. The edge of the valence and conduction bands was read from the DOS in Fig. \ref{dos}. The interval between the calculated data points of the DOS is about 0.04 eV. The Fermi-Dirac distribution decays so fast that its value may already decrease from 0.5 at the Fermi level to 0.2 at the closest data point, and most of the carrier density is contributed from the closest three data points near the Fermi level on one side. If visibly the DOS increases linearly from the edge of band for the first five data points, we linear-fit these five points to obtain an analytical form of the DOS and integrate over this small range. Otherwise, we sum up areas of rectangles from the edge of band to the other end. The uncertainty on the carrier density from this is about 10\%. There is a further uncertainty due to some of the interstitial states between graphene and Si being wrongly projected to Si, which could modify the presented values of carrier density by a factor of 2 at most. We therefore only focus on the changes by orders of magnitude.

The Fermi level of the unsupported graphene is at the Dirac point (despite the in-plane strain), as expected. The carrier densities of electrons and holes are the same and the values are consistent with those reported in the literature.\cite{Fang07,Wang16,Yin14} The in-plane strain does not affect the carrier density much, which is reasonable as the states near the Dirac point are all from the out-of-plane $\pi$-electronic orbitals. When the graphene is placed on the Si (111) surface at the vdW distance, the Fermi level is shifted by 0.7 eV into the conduction band from the Dirac point, resulting in an increase of the electron density by more than two orders of magnitude. When a covalent bond is formed to the Si (111) surface, the Fermi level is shifted slightly less into the conduction band (by 0.6 eV), but to an energy range where the graphene DOS is large, resulting in an increase of carrier density by three orders of magnitude.

The significant increase in the charge density (and the work function from the previous section) is due to an unconventional $n$-type doping, which is due not to charge transfer, but to a modification of the DOS. The shift of the Fermi level into the conduction band is unexpected, as the Fermi level of the system (graphene placed on Si) is higher than both the Dirac point of graphene and the top of the valence band of Si. A possible explanation is that the number of available states of graphene and/or Si below the Dirac point of graphene has decreased, due to the interaction between the graphene and Si surface.

On Si (100) and (110) surfaces, the in-plane strain is anisotropic to achieve lattice match. Band gaps appear as a result of the strain, which is of no interest in this paper. The Fermi levels in all the modelled cases on (100) and (110) Si surfaces are at the top of the valence band of graphene. So graphene is undoped on Si (100) and (110) surfaces. In most cases, the carrier densities remain similar to that of pristine graphene, except when the graphene is placed on Si (100) at the vdW distance. Here the DOS increases sharply near the Dirac point (Fig. \ref{dos} (f)), which results in an increase of the carrier density by a factor of 80.

The larger the Si surface density, the more robust the surface, and the more it can modify the $\pi$-orbitals and the DOS of graphene. The carrier density of graphene on Si (100) is therefore greatly increased, and it can be further increased on Si (111) (with the highest surface density), where the Fermi level is shifted through the modification of the DOS.

\section{Conclusions}
In this paper, we modelled monolayer graphene on Si substrates having three different surface orientations, (111), (100) and (110). We found that the role of the substrate is remarkably varied according to its orientation. Even when the interaction with the graphene is only vdW, there can be large modifications on the graphene DOS, leading to large effects (work function and carrier density). And the formation of Si-C covalent bonds is also highly orientation-dependent.   
\begin{acknowledgments}
YWS is grateful for the valuable discussion with Dr Dimitrios Papageorgiou, Dr Jan Mol from Queen Mary University of London, and Dr Wei Liu from Zhejiang University. YWS, OF and CJH are grateful for the financial support from the Innovate UK [Project No. 104714]. OF is a Royal Society University Research Fellow (UF140372; URF\text{\textbackslash}R\text{\textbackslash}201013). The computational results presented have been achieved [in part] using the Vienna Scientific Cluster (VSC).
\end{acknowledgments}


\bibliography{apssamp1}

\end{document}